\documentclass[prl,superscriptaddress,showpacs,twocolumn,amsfonts,amsmath]{revtex4}
\usepackage[final]{graphicx}

\newlength\figurewidth
\AtBeginDocument{\setlength\figurewidth{.9\linewidth}}

\usepackage{color}
\let\omp\marginpar\relax\def\marginpar#1{\omp{\color{red}#1}}

\newcommand{\cv}{{\bf c}}
\newcommand{\xv}{{\bf x}}
\newcommand{\uv}{{\bf u}}
\newcommand{\vv}{{\bf v}}

\begin{document}
\title{Lattice Boltzmann versus Molecular Dynamics simulation of nano--hydrodynamic flows}
\date{\today}
\def\mainz{%
  \affiliation{Institut f\"ur Physik, Johannes-Gutenberg-Universit\"at Mainz,
   Staudinger Weg 7, D-55099 Mainz, Germany}}
\def\roma{%
  \affiliation{Istituto Applicazioni Calcolo, CNR, V.le del Policlinico 137, 00161, Roma, Italy}}
\author{J\"urgen Horbach}\mainz
\author{Sauro Succi}\roma

%
\begin{abstract}
A fluid flow in a simple dense liquid, passing an obstacle in a
two--dimensional thin film geometry, is simulated by Molecular Dynamics
(MD) computer simulation and compared to results of Lattice Boltzmann
(LB) simulations.  By the appropriate mapping of length and time units
from LB to MD, the velocity field as obtained from MD is quantitatively
reproduced by LB.  The implications of this finding for prospective
LB-MD multiscale applications are discussed.
\end{abstract}

\maketitle

Micro and nano--hydrodynamic flows are playing an increasing role for many
applications in material science, chemistry, and biology \cite{MICRO}.
To date, the leading tool for the numerical investigation of nano
and micro--hydrodynamic flows is provided by molecular dynamics (MD)
simulations \cite{MD}. In principle, MD yields a correct description of
fluids on microscopic and hydrodynamic scales.  But typical length and
time scales that can be covered by MD simulations of dense liquids are of
the order of a few tens of nanometers and about a few hundred nanoseconds,
respectively. As a result, the quest for cheaper --- and yet physically
realistic alternatives --- is a relentless one.  Mesoscopic models,
and notably those arising from kinetic theory, are natural candidates
to fill this gap because they operate precisely at intermediate scales
between the atomistic and continuum levels. Very recently, the so--called
Lattice--Boltzmann (LB) method, where a kinetic equation is solved on
a lattice, has proven successful to model characteristic features of
microscopic flows, such as the occurrence of slip boundary conditions
near solid walls \cite{SLIPMOD}.

However, whether LB is also able to reproduce complex nanoscopic flows
on a quantitative level, in particular for a dense liquid, remains
an open issue. Although, pioneering work of various authors has shown
that hydrodynamics holds nearly down to the molecular scale in simple
dense liquids (see, e.g.~\cite{HYDMOL}), it is not obvious whether
in the vicinity of wall--fluid or obstacle--fluid boundaries at least
pair interactions on an atomistic level have to be taken into account
to yield a quantitative description of fluid flows. Although LB is a
particle--based method, it cannot resolve the short ranged structural
order of a dense liquid since it describes a structureless lattice gas
with no many--body interactions. Thus, it is not clear whether one can
match the LB method with MD simulations of complex flow structures in
dense liquids (e.g.~coherent nanostructures).

In this paper, we aim to establish a quantitative mapping between the
resolution requirements of a LB versus MD simulation of a non--trivial
nano--hydrodynamic flow.  Far from being a purely academic exercise, this
is a primary issue to set a solid stage for future multiscale applications
combining LB with atomistic methods.  To this end, we have performed MD
simulations of a two--dimensional Lennard--Jones fluid confined between
walls that passes a thin plate--like obstacle while subject to a gravitational
force field. This system is then modeled by the LB method. We demonstrate
that, after the appropriate mapping of length and time units, there is
a quantitative agreement between the flow fields computed with MD and LB.

First, we give a brief introduction to the LB method and the LB model
used in this work. The LB method as a mesoscopic simulation tool has met
with significant success in the last decade for the simulation of many
complex flows \cite{LBE}.  We consider the LB equation in its original
matrix form \cite{HSB}:
\begin{eqnarray}
f_i(\xv+\Delta t \cv_i,t+\Delta t)-f_i(\xv,t) & = & \nonumber \\
-\sum_j \Omega_{ij} \Delta t && \! \! \! \! \! \! \! \! \! \! 
[f_j(\xv,t) -f_j^{\rm e}(\xv,t)]
\label{BGK_eq}
\end{eqnarray}
where $\Delta t$ is the time unit and $f_i(\xv,t) \equiv
f(\xv,\vv=\cv_i,t)$, $i=0, \ldots n$, is the probability of finding a
particle at lattice site $\xv$ at time $t$, moving along the lattice
direction defined by the discrete speed $\cv_i$.  The left--hand
side of Eq.~(\ref{BGK_eq}) represents the molecular free--streaming,
whereas the right--hand side represents molecular collisions via a
multiple--time relaxation towards local equilibrium $f_i^{\rm e}$
(a local Maxwellian expanded to second order in the fluid speed).
The leading non--zero eigenvalue of the collision matrix $\Omega_{ij}$
fixes the fluid kinematic viscosity as $\nu=c_{\rm s}^2(1/\omega-1/2)$
(in lattice units $\Delta t=\Delta x=1$), where $c_{\rm s}$ is the
sound--speed of the lattice fluid, $1/\sqrt{3}$ in the present work.
In order to recover fluid dynamic behaviour at macroscopic scales,
the set of discrete speeds must be chosen in such a way as to conserve
mass and momentum at each lattice site.  Once these conservation
laws are fulfilled, the fluid density $\rho=\sum_i f_i$, and speed
$\uv = \sum_i f_i \cv_i/\rho$ can be shown to evolve according to the
quasi--incompressible Navier--Stokes equations of fluid--dynamics.

In the bulk flow, LB is essentially an efficient Navier-Stokes solver in
disguise. The major advantages over a purely hydrodynamic description
are: i) the fluid pressure is locally available site--by--site,
with no need of solving a computationally demanding Poisson problem,
ii) momentum diffusivity is not represented by second order spatial
derivatives, but it {\it emerges} instead from the first--order
LB relaxation--propagation dynamics. As a result, the time--step
scales only linearly (rather than quadratically) with the mesh size,
which is an important plus for down--coupling to atomistic methods,
iii) highly irregular boundaries can be handled with ease because
particles move along straight trajectories. This contrasts with the
hydrodynamic representation, in which the fluid momentum is transported
along complex space--time dependent trajectories defined by the flow
velocity.  

At the fluid--solid interface, we adopt the no--slip boundary
condition, $\vec{u}=0$.  This is imposed by the standard bounce--back
procedure, i.e.~by reflecting the outgoing populations back into
the fluid domain along their specular direction.  With reference
to particles propagating south--east($\searrow$) from a north--wall
boundary placed at $z=H+1/2$, the bounce-back rule simply reads as
$f_{\searrow}(x,y,H+1)=f_{\nwarrow}(x+1,y,H)$ where the lattice spacing
is made unity for convenience. This corresponds to a stylized two--body
hard--sphere repulsion, with interaction range equal to $\sqrt{2}$
lattice units.

In this work, we use a standard nine--speed D2Q9 model \cite{LBE} to
simulate a two--dimensional channel flow of size $L$ and $H$ along the
streamwise and cross--flow directions, respectively.  At a distance $L/4$
from the inlet, a thin flat plate of height $h$ is placed, perpendicular
to the direction of the main flow.  The fluid is driven by a constant
body force $f_x$ in $x$--direction which, in the absence of the vertical
plate, would produce a parabolic Poiseuille profile of central speed
$U=f_xH^2/(8\nu)$, $\nu$ being the kinematic viscosity of the fluid.

A similar two--dimensional channel flow of a fluid is considered in the
MD simulations.  As a model for the interactions between the particles,
a Lennard--Jones potential is used that is truncated at its minimum and
then shifted to zero. It has the following form:
\begin{equation}
V(r) = \left\{
    \begin{array}{ll}
    4 \epsilon \left[ \left(\frac{\sigma}{r}\right)^{12} 
     - \left(\frac{\sigma}{r} \right)^{6} \right] + 
       \epsilon & \mbox{for $r<2^{1/6} \sigma$} \\
    0      & \mbox{otherwise,}
    \end{array}
  \right. 
\end{equation}
where $r$ denotes the distance between two particles. The units are chosen
such that $\epsilon$, $\sigma$, the Boltzmann constant $k_{\rm B}$ and
the mass of the particles $m$ are set to one. The particles are confined
into a rectangular box of size $L \times H$ with $L=200 \sigma$ and $H=106
\sigma$. Moreover, an obstacle is placed along a line at $x=L/4$. This
obstacle consists of $41$ fixed particles between $y=33.0 \sigma$
and $y=73.0\sigma$ with an equidistant spacing of $1 \sigma$. Thus,
the obstacle has an effective width of about $1 \sigma$ and a height
$h=40 \sigma$. For the density we chose $n=N/(LH)=0.8\sigma^{-3}$
corresponding to $N=16879$ particles in the simulation box. The
equations of motions were integrated using the velocity form of the
Verlet algorithm with a time step of $dt_{\rm md}=0.01 \tau_{\rm md}$
with $\tau_{\rm md}\equiv\sqrt{m \sigma^2/(48 \epsilon)}$. In order to
keep the temperature constant, a Nos\'e--Hoover thermostat was applied
\cite{binder96}.

The MD simulations were all done at temperature $T=5.3 \epsilon/k_{\rm
B}$.  Ten independent runs were performed in order to obtain
reasonable statistics. First, the systems were equilibrated for
30000 time steps. Then, walls were introduced in the system by
giving all the particles at $x<3.0\sigma$ and at $x>103.0\sigma$ zero
velocity. The choice of these rough walls provides the absence of any
layering effects of the fluid near the walls. The system with walls,
i.e.~with the immobilized particles, was then equilibrated for another
20000 time steps, followed by runs over two million time steps with a
gravitational force field perpendicular to the walls ($x$ direction)
with magnitude $f_x=2\times 10^{-4}$ at each particle.  Note that in
the MD with external force field the Nos\'e--Hoover thermostat was
only applied in $y$-direction, i.e.~perpendicular to the flow field.
Results were collected after $2 \times 10^5$ time steps when the steady
state was clearly reached. Then, every 1000 steps a configuration with
positions and velocities of the particles was stored from which all the
quantities that are presented in the following were computed. In total
the average was over 18000 configurations.

In addition to the latter runs, also MD simulations were performed for
a system without obstacle. Here, the aim was to determine the kinematic
viscosity from the Poiseuille profile that forms at steady state. As
a result, we obtained the value $\nu=0.5$ for the kinematic viscosity
(in MD units). Thus, with $U_{\rm f}=0.07$ the average flow velocity
in the simulations with obstacle, the Reynolds number is $Re=U_{\rm
f}h/\nu\approx5.6$, above the critical value for the onset of coherent
structures.

In order to compare the results of the MD with those of LB, a conversion
of MD units into LB units is required.  Space and velocity (time)
conversion proceed as follows: $L=\tilde L_{\rm lb} \Delta x = \tilde
L_{\rm md} \sigma$, $U=\tilde U_{\rm lb} \Delta x/\Delta t = \tilde
L_{\rm md} \sigma/\tau_{\rm md}$.  This yields $\tilde L_{\rm lb} =
\tilde L_{\rm md} \sigma/\Delta x$ and $\tilde U_{\rm lb} = \tilde
U_{\rm md} \frac{\sigma}{\Delta x} \frac{\Delta t}{\tau_{\rm md}}$.
The conversion of the kinematic viscosity, central to the definition
of the Reynolds number, is given by $\tilde \nu_{\rm lb} = \tilde
U_{\rm md} (\frac{\sigma}{\Delta x})^2 \frac{\Delta t}{\tau_{\rm md}}$.

Reducing $\Delta x$ to values of the order of $\sigma$ means that the LB
simulation would resolve the structure of the fluid, if this happened
to be included into the lattice kinetic equation. This is not the case
here. The interesting question, though, is whether the absence of these
structural effects does hamper the correct reproduction of bulk--flow
features.  This is very similar, although on totally different scales, to
the issue of subgrid scale modeling of turbulent flows: do the unresolved
scales spoil the physics of the resolved ones?

In order to map LB units onto MD units, we proceeded as follows: First,
we identified the LB mesh--spacing with a fraction or a multiple of
$\sigma$. Then, the time conversion factor was determined from the
kinematic viscosity, such that the same Reynolds number, $Re\approx 5.6$,
was yielded in LB and MD. In this way, LB simulations were done for
$\Delta x=0.25\,\sigma$, $0.5\,\sigma$, $1.0\,\sigma$, and $2.0\sigma$
which respectively corresponds to the values $\tilde{\nu}_{\rm lb}=1.92,
0.96, 0.48,$ and 0.24. The LB simulation ran over 10000 steps, which was
found sufficient to keep time changes of the overall velocity profile
at least within third digit accuracy. The CPU time required for the LB
simulations varied between about 1\,min to 45\,min on a Pentium 4 with
2.8 GB clock rate, depending on the chosen resolution. This has to be
compared to the total computational load for the MD simulation, which
was about 1 week on 10 Pentium 4 processors.

\begin{figure}
\centering
\includegraphics*[width=.3\textwidth]{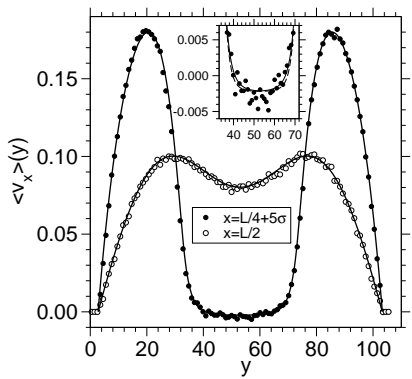}
\caption{\label{fig1}
Steady--state streamwise velocity profiles $u(y)=<v_x(y)>$ as a function
of the crossflow coordinate $y$ at $x=L/4+5\sigma$ (i.e., $5\sigma$ behind
the obstacle) and $x=L/2$ as obtained from the MD as indicated. 
The solid lines are the corresponding results from the LB with resolution
$\Delta x=0.25\sigma$. Both the MD and LB curves are averaged along the $x$ direction
over a region of $\pm 3\sigma$. The inset shows a magnification of the
region around the minimum for the profile at $x=L/4+5\sigma$ (here, the 
dashed line is the LB result without the averaging along the $x$ direction.
}
\end{figure}
In Fig.~\ref{fig1}, we show the steady--state streamwise velocity profile
$u(y)\equiv <v_x(y)>$ as a function of the crossflow coordinate $y$ at
$x=L/4+5\sigma$ and $x=L/2$, for the LB with $\Delta x=0.25\sigma$, as
compared to the corresponding profile obtained with the MD simulations.
Here, both MD and LB data are averaged along the $x$ direction over a region
of $\pm 3\sigma$. This leads to a smoothening of the MD data, but,
as the LB results show, has only minor effects on the profiles (this
can be infered from the dashed line in the inset of Fig.~\ref{fig1}
which shows the LB data without averaging along the $x$ direction). From the
data, very good agreement is observed between MD and LB results. Of
particular interest is the region around the minimum in the curve for
$x=L/4+5\sigma$, where the velocity field $<v_x>$ becomes negative.
These negative velocities are due to the formation of vortices behind
the obstacle. As the inset of Fig.~\ref{fig1} shows, even this region
of negative velocities is well reproduced by the LB simulation.

\begin{figure}
\centering
\includegraphics*[width=.3\textwidth]{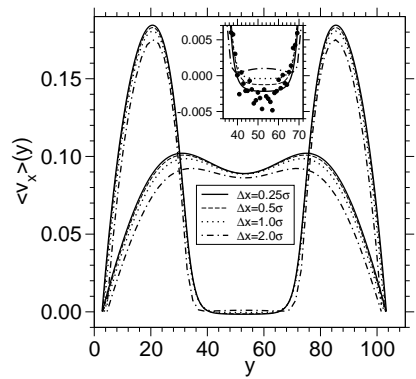}
\caption{\label{fig2}
Steady--state streamwise velocity profile $u(y)=<v_x(y)>$ as a function
of the crossflow coordinate $y$ at $x=L/4+5\sigma$ and $x=L/2$ for
the LB simulations with different choices of $\Delta x$ as indicated.
As in Fig.~\ref{fig1}, the inset shows a magnification of the region
around the minimum in the curves for $x=L/4+5\sigma$.
}
\end{figure}
In Fig.~\ref{fig2} the same quantities as in Fig.~\ref{fig1} are shown
for the LB runs with different choices of $\Delta x$.  From this figure,
it is apparent that lack of resolution of the atomic scale $\sigma$
generates significant departures of the LB results from the MD data.
This is an informative result, for it implies that microscopic length
scales, {\it although absent from the LB description}, must nonetheless be
resolved if the flow profile away from the walls is to be quantitatively
captured by the coarse--grained simulation. This is especially evident
in the region where vortices form, i.e.~just behind the obstacle. In
this region, a quantitative agreement with the MD requires a resolution
as high as $\Delta x=0.25\sigma$ (see the inset of Fig.~\ref{fig2}).

\begin{figure}
\centering
\includegraphics[width=0.3\textwidth]{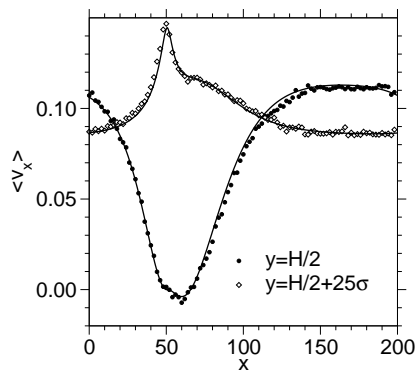}
\caption{\label{fig3}
Steady--state streamwise velocity profile $u(y)=<v_x(y)>$ as a function
of $x$ at $y=H/2$ and $y=H/2\pm25\sigma$. The symbols show the MD data 
as indicated. The solid lines are the corresponding LB data with 
$\Delta x=0.25\sigma$. 
}
\end{figure}
\begin{figure}
\centering
\includegraphics*[width=0.39\textwidth]{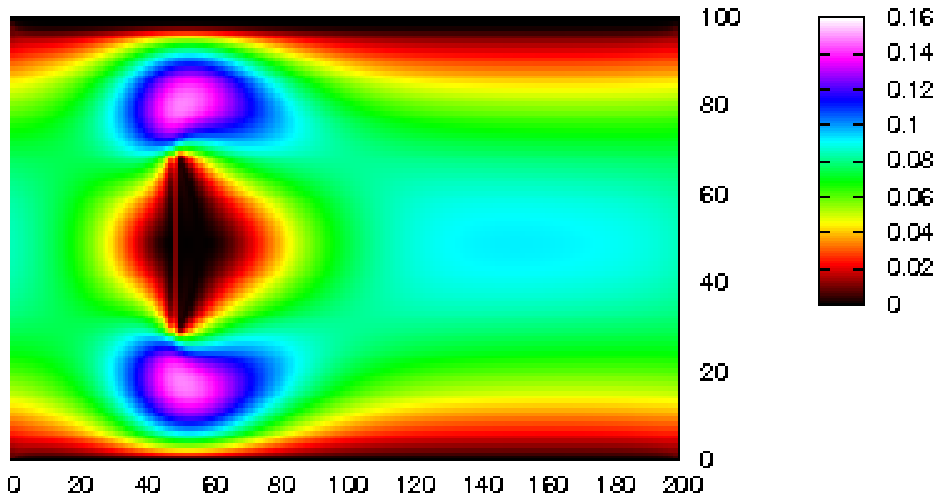}
\includegraphics*[width=0.38\textwidth]{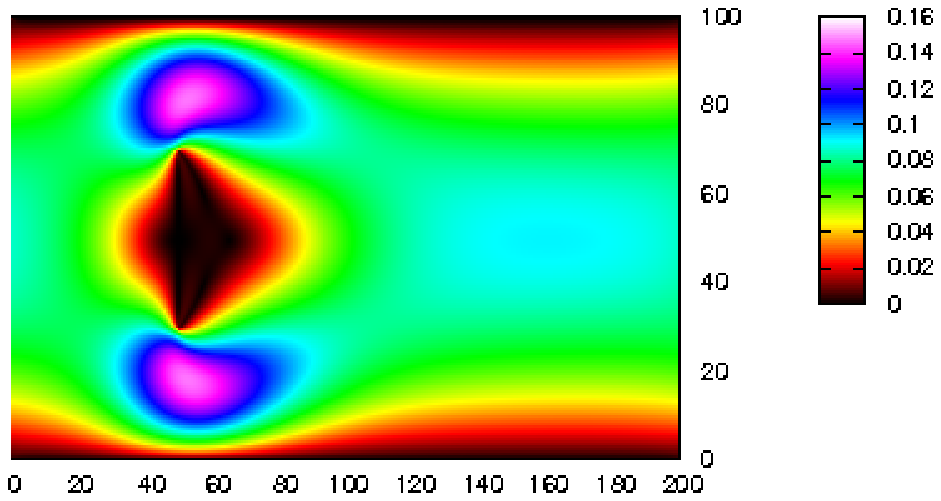}
\caption{\label{fig4}
Upper panel: The color map shows the magnitude of the velocity field,
$|\vec{u}(x,y)|=\sqrt{<v_x>^2+<v_y>^2}$ at steady state as obtained from the MD
simulation. Lower panel: The same result as obtained from the LB simulation
with $\Delta x=0.25 \sigma$ is shown.}
\end{figure}
In Fig.~\ref{fig3} the streaming velocity $<v_x>$ in the direction
parallel to the flow is considered for $y=H/2$ and $y=H/2 +25 \sigma$.
In this case LB and MD data were averaged over a region of $\pm 2
\sigma$.  Note that in the MD case, also the data at $y=H/2-25 \sigma$
were used for the average of the profile at $y=H/2 +25 \sigma$, since
the profile is expected to be symmetrical with respect to the central
line at $y=H/2$. The comparison between LB and MD shows again a very good
agreement. This is not a foregone result,  since there are non--trivial
features in the vicinity of the obstacle, i.e.~around $x=50\sigma$,
such as the shoulder at $x=55\sigma$ in the curve for $y=H/2 +25
\sigma$.  Fig.~\ref{fig4} displays the magnitude of the velocity field,
$|\vec{u}(x,y)|=\sqrt{<v_x>^2+<v_y>^2}$, as obtained from MD (upper plot)
and LB (lower plot). Here, it is illustrated that the whole velocity
field is quantitatively reproduced.

In conclusion, we have demonstrated an example of a complex
nanoscopic fluid flow, in which  a spatial handshaking between LB and MD is possible.
The mapping of LB onto MD requires a conversion of
time and length units from LB to MD and one has to choose an appropriate
grid resolution for the LB fluid. For the dense fluid considered in the
MD simulation of this work, a grid resolution $\Delta x =0.25\sigma$
is required to reproduce the flow features quantitatively. 
Remarkably, non--hydrodynamic (finite-Knudsen) effects appear to be silent, at least at
steady state. This is likely to be a benefit of the 
matrix formulation of the collision operator \cite{HSB,MTR}, as 
compared to the more popular single-time relaxation form.

The present results indicate that there appears to be a sound ground
for prospective multiscale applications based on the combined use of
(multigrid) LB \cite{MULBE} with MD.  For instance, by using multigrid
LB with, say, $6$ levels of resolution, one could couple LB with MD
at the finest scale $\Delta x_{\rm fine} \sim \sigma$, typically near
the boundaries, and then progressively increase the LB mesh--size so
as to reach a hundred--fold larger mesh spacing $\Delta x_{\rm bulk}=
2^6 \Delta x_{\rm fine}$ in the bulk flow.  On--the--fly time--coupling
is obviously more demanding.  In fact, our simulations indicate that
even with the finest resolution $\Delta x = \sigma/4$, LB is about a
factor $2000$ faster than MD (see above). The most direct
strategy for 'on--the--fly' LB--MD coupling is to apply LB everywhere
in the fluid domain and leave MD only in small portions, typically in a
ratio $1:1000$ to the global domain.  Moreover, this factor $1000$ gap
could be partially bridged by resorting to very recent time--adaptive LB
procedures \cite{RASIN}.  This indicates that, although very demanding,
even time--coupling between LB and MD may become soon feasible for the
numerical investigation of complex nanoflows.

Of course, several open issues remain.  Our comparisons were done
at steady state and it is not clear to what extent transient states
are correctly described by LB.  Moreover, it will be interesting to see
whether the LB method still yields a quantitative description in the case
of slip--flow and other complex phenomena at the fluid--solid interface.
These issues make interesting subjects for forthcoming studies.

\begin{acknowledgments}
We are grateful to Kurt Binder for very valuable discussions.  One of
the authors (S. S.) thanks the Alexander von Humboldt Foundation for
financial support and Kurt Binder and his group for kind hospitality
and the stimulating atmosphere during his stays.
\end{acknowledgments}

\end{document}